\documentclass{article} 

\usepackage[numbers]{natbib}       
\usepackage{float}                 

\usepackage{amsmath}              
\usepackage{algorithm}            
\usepackage{algorithmic}          

\usepackage{multirow}
\usepackage{booktabs}             
\usepackage{nicefrac}             
\usepackage{microtype}            
\usepackage{lipsum}               

\usepackage[utf8]{inputenc}       
\usepackage[T1]{fontenc}          
\usepackage{amsfonts}             
\usepackage{float}
\usepackage{hyperref}             
\usepackage{url}                  

\usepackage{graphicx}
\graphicspath{{./images/}}        

\title{SER-Diff: Synthetic Error Replay Diffusion for Incremental Brain Tumor Segmentation}

\author{Sashank Makanaboyina\\
DePaul University, Chicago USA\\
\texttt{smakanab@depaul.edu}}

\date{} 

\begin{document}
\maketitle

\begin{abstract}
Incremental brain tumor segmentation is critical for models that must adapt to evolving clinical datasets without retraining on all prior data. However, catastrophic forgetting, where models lose previously acquired knowledge, remains a major obstacle. Recent incremental learning frameworks with knowledge distillation partially mitigate forgetting but rely heavily on generative replay or auxiliary storage. Meanwhile, diffusion models have proven effective for refining tumor segmentations, but have not been explored in incremental learning contexts. We propose Synthetic Error Replay Diffusion (SER-Diff), the first framework that unifies diffusion-based refinement with incremental learning. SER-Diff leverages a frozen teacher diffusion model to generate synthetic error maps from past tasks, which are replayed during training on new tasks. A dual-loss formulation combining Dice loss for new data and knowledge distillation loss for replayed errors ensures both adaptability and retention. Experiments on BraTS2020, BraTS2021, and BraTS2023 demonstrate that SER-Diff consistently outperforms prior methods. It achieves the highest Dice scores of 95.8\%, 94.9\%, and 94.6\%, along with the lowest HD95 values of 4.4 mm, 4.7 mm, and 4.9 mm, respectively. These results indicate that SER-Diff not only mitigates catastrophic forgetting but also delivers more accurate and anatomically coherent segmentations across evolving datasets.
\end{abstract}

\textbf{keywords}:{Incremental Learning, Catastrophic Forgetting, Knowledge Distillation, Diffusion Models, Brain Tumor Segmentation, Medical Image Analysis}
 
\section{Introduction}
\label{sec:intro}
Brain tumor segmentation from MRI data is a cornerstone of neuro-oncology, providing essential support for accurate diagnosis, treatment planning, and longitudinal patient monitoring \cite{Bauer2013Survey}. Among deep learning solutions, U-Net \cite{re1} and its three-dimensional extension, 3D U-Net \cite{re2}, have become the standard architectures, delivering strong performance within supervised learning paradigms. However, these models typically require large, annotated datasets and significant computational resources \cite{re1}, which restricts their scalability in clinical practice, where annotations are scarce and imaging protocols vary between institutions. To address these limitations, recent studies have investigated generative modeling strategies. In particular, Denoising Diffusion Probabilistic Models (DDPMs) have gained traction for cross-modal MRI synthesis, enabling the reconstruction of missing modalities from available contrasts \cite{re2}. For instance, DDMM Synth \cite{DDMM} employs diffusion processes to generate the T1ce modality with high fidelity based on other MRI sequences.

Precise brain tumor segmentation from MRI scans plays a pivotal role in clinical diagnosis and treatment planning. Deep learning methods have significantly advanced this field by achieving high accuracy across diverse benchmarks. However, in realistic clinical environments, models must adapt to continuously evolving data distributions, such as new patients, scanners, and tumor types, without access to all prior data. Retraining from scratch is computationally expensive and impractical due to privacy restrictions, motivating the need for Incremental Learning (IL) approaches \cite{IL2025}. 

Despite their promise, IL methods often suffer from catastrophic forgetting, where performance on previous tasks degrades as new tasks are learned. Prior work has explored regularization-based strategies, generative replay, and knowledge distillation, but these approaches either fail to fully preserve structural information or incur high computational costs.

Diffusion models were initially introduced for image generation, where they proved capable of producing diverse and high-quality samples \cite{Ho2020}. Their utility, however, extends well beyond generative tasks, as they have been adapted to a wide range of computer vision problems, including image editing \cite{re7}, super-resolution \cite{re8}, and medical image segmentation \cite{re9}. These advances emphasize the flexibility and robustness of diffusion-based methods across different application domains. The emergence of large-scale frameworks such as DALL·E 2 \cite{re10}, Imagen \cite{re11}, and Stable Diffusion \cite{re12} has further pushed the boundaries of the field, delivering photorealistic and semantically coherent images \cite{re14}. Collectively, these developments illustrate the expanding impact of diffusion models in both general-purpose and specialized vision tasks. Recent work has highlighted the importance of diffusion models for refining segmentation predictions. For instance, in \cite{DMCIE2025}, we proposed DMCIE, which integrates segmentation error maps with multi-modal inputs through diffusion, significantly improving boundary accuracy. Building on this idea, ReCoSeg \cite{shortmidl} introduced residual-guided cross-modal synthesis for brain tumor segmentation, demonstrating the utility of synthetic T1ce generation for enhancing performance. More recently, ReCoSeg++ \cite{ReCoSeg2025} extended this framework with multi-stage residual-guided refinement, further improving segmentation on heterogeneous MRI data.

Although diffusion models have shown exceptional capabilities for image generation and medical segmentation, but their potential has not yet been fully explored in the incremental learning setting. Standard generative replay strategies synthesize entire images, which are computationally demanding and not optimized for correcting segmentation errors. To address this gap, we propose a novel framework, Synthetic Error Replay Diffusion (SER-Diff), which combines the adaptability of incremental learning with the refinement power of diffusion models. Unlike prior generative replay approaches, SER-Diff leverages synthetic error maps generated by a frozen teacher diffusion model to retain past knowledge. This design reduces computational overhead and enables targeted error correction, ensuring both adaptability to new tasks and retention of previously learned information. Our work introduces SER-Diff as the first diffusion-based incremental learning framework for brain tumor segmentation. By unifying error-guided diffusion with synthetic replay, SER-Diff provides a principled approach to mitigating catastrophic forgetting while improving segmentation accuracy across evolving clinical datasets.

\section{Related work}
\label{sec:rel}
Magnetic Resonance Imaging (MRI) has become a cornerstone of modern radiology due to its superior soft-tissue contrast and ability to acquire images in multiple planes and modalities \cite{reMRI}. Unlike Computed Tomography (CT), which relies on ionizing radiation, MRI uses strong magnetic fields combined with radio-frequency pulses to produce detailed views of brain structures. Multi-sequence imaging, such as T1-weighted (T1w), T2-weighted (T2w), Fluid-Attenuated Inversion Recovery (FLAIR), and Diffusion-Weighted Imaging (DWI), provides complementary information that is critical for brain tumor detection and characterization \cite{reAu}. Deep learning methods have shown remarkable progress in medical image segmentation. Beyond medical imaging, deep learning has also demonstrated effectiveness in broader IoT and heart diseases domains \cite{heart}.

In brain tumor analysis, Convolutional Neural Networks (CNNs) remain the most widely adopted architectures. Models such as U-Net and its 3D extensions \cite{re2} have achieved strong performance but can be less effective when facing irregular tumor shapes or heterogeneous lesion appearances. To overcome these limitations, attention-based architectures such as TransBTS and CANet have been proposed, enhancing contextual feature learning. However, these approaches can still produce incomplete or blurred boundaries.

Diffusion-based models have recently emerged as powerful alternatives for segmentation refinement. By iteratively modeling noise distributions, Denoising Diffusion Probabilistic Models (DDPMs) have demonstrated effectiveness in generating realistic structures and correcting segmentation errors. Recent works have explored their application in brain tumor segmentation, including residual-guided cross-modal approaches such as ReCoSeg \cite{shortmidl} and its extended version ReCoSeg++ \cite{ReCoSeg2025}, which leverage synthesized T1ce contrasts to enhance tumor delineation on BraTS datasets. These methods underscore the growing role of diffusion-based refinement in advancing segmentation accuracy.

Incremental Learning (IL) has also attracted increasing attention in medical imaging, aiming to adapt models to evolving datasets without retraining on all prior data. IL strategies are typically categorized into rehearsal-based, regularization-based, and architectural methods. Rehearsal methods require storing past samples, raising privacy and storage concerns, while architectural expansions increase model complexity. Knowledge distillation (KD) is a widely adopted solution, though it often struggles to preserve fine-grained spatial details. Recent work \cite{IL2025} has demonstrated that IL can benefit from generative replay and KD, but most approaches remain limited when applied to segmentation tasks.

While diffusion models have proven highly effective for segmentation refinement, no prior work has integrated them into the incremental learning setting. Our proposed framework, SER-Diff, addresses this gap by replaying synthetic error maps generated by a frozen teacher diffusion model, combining the strengths of generative replay with targeted error-guided refinement.

\section{Methodology}

The proposed framework, SER-Diff, consists of three core components. First, we introduce the synthetic error replay mechanism (\autoref{sec:ser}), where a frozen teacher diffusion model generates error maps from prior tasks that are replayed during incremental training. Second, we describe the diffusion-based refinement process (\autoref{sec:refine}), which leverages these error maps to guide denoising and improve segmentation consistency across tasks. Finally, we present the dual-loss training strategy (\autoref{sec:dualloss}), which combines Dice loss for learning new tasks with knowledge distillation loss for replayed errors, ensuring both adaptability to new data and retention of previously acquired knowledge.

\subsection{Synthetic Error Replay}
\label{sec:ser}
To avoid storing raw data from past tasks, we propose a synthetic error replay mechanism. A teacher diffusion model $T$ is trained on the first task and then frozen. For each new task $D_k$, $T$ generates synthetic error maps $\hat{E}_{old}$ conditioned on multimodal MRI inputs $c$ (e.g., T1, T1ce, T2, FLAIR). These maps represent the discrepancies between predictions and ground truth that the teacher would have corrected in earlier tasks. By replaying these synthetic errors, the student model gains access to compact task-specific knowledge without violating privacy constraints or requiring large storage resources.

\subsection{Diffusion-Based Refinement}
\label{sec:refine}
The student model $S$ is trained incrementally using both the current task data and the synthetic error maps. The refinement process follows a diffusion-based denoising formulation. At each time step $t$, noisy error maps are sampled as
\[
q(x_t \mid x_{t-1}) = \mathcal{N}\left(\sqrt{1-\beta_t}x_{t-1}, \beta_t I\right),
\]
and the reverse denoising step is parameterized by the student as
\[
p_\theta(x_{t-1} \mid x_t, c) = \mathcal{N}\left(\mu_\theta(x_t, t, c), \Sigma_\theta(x_t, t, c)\right),
\]
where $c$ is the multimodal MRI input. By conditioning on both current-task MRI sequences and synthetic error maps, the diffusion process guides the student model to iteratively correct tumor boundaries and improve segmentation consistency across tasks.

\subsection{Dual-Loss Training Strategy}
\label{sec:dualloss}
To ensure adaptability to new tasks while retaining prior knowledge, SER-Diff employs a dual-loss training objective:
\[
\mathcal{L} = \underbrace{\text{Dice}(M_S, M_{GT}) + \text{BCE}(M_S, M_{GT})}_{\text{Current Task Loss}} 
+ \lambda \underbrace{\Big( \|f_S(\hat{E}_{old}) - f_T(\hat{E}_{old})\|_2^2 + \text{Cov}(Z_S, Z_T) \Big)}_{\text{Knowledge Distillation Loss}}.
\]
The first component ensures voxel-level segmentation accuracy on the current task $D_k$, while the second aligns the student’s feature space with that of the frozen teacher using synthetic error maps. The covariance regularization further stabilizes feature embeddings and prevents degradation of previously learned representations. This joint optimization allows SER-Diff to achieve high accuracy on new tasks while mitigating catastrophic forgetting of earlier tasks.

\section{Experimental Setup}

Experiments are conducted on two benchmark datasets for brain tumor segmentation. The BraTS2020 dataset is used as the initial training task and consists of 369 subjects with four MRI modalities for each case, including T1, T1ce, T2, and FLAIR. All volumes are co-registered, skull-stripped, and resampled to $1 \times 1 \times 1$ mm resolution. For incremental adaptation, we employ the BraTS2021 and BraTS2023 datasets, which introduces additional heterogeneity due to scanner variations and broader patient populations. This sequential configuration reflects a realistic continual learning scenario in which models must adapt to new data distributions without revisiting samples from earlier tasks.

To evaluate the effectiveness of SER-Diff, we compare it against several strong baselines. Standard incremental learning with knowledge distillation (KD) is included as a widely adopted approach where a student network is updated on new tasks under the guidance of a frozen teacher. In addition, we consider DMCIE \cite{DMCIE2025}, a diffusion-based refinement strategy that improves segmentation accuracy but does not explicitly address incremental learning. Finally, we include CorrDiff \cite{CorrDiff7}, a state-of-the-art diffusion model for systematic error correction that is trained in a non-incremental setting, serving as a reference upper bound.

Evaluation is performed using two complementary metrics. The Dice Similarity Coefficient (DSC) measures volumetric overlap between predicted tumor masks and ground truth labels, providing a primary indicator of segmentation accuracy. To quantify knowledge retention across tasks, we compute the Forgetting Rate (FR), defined as
\[
FR = \frac{1}{K-1} \sum_{i=1}^{K-1} \left(\max_{t} Acc_{i}^{t} - Acc_{i}^{K}\right),
\]
where $Acc_{i}^{t}$ denotes performance on task $i$ after training at stage $t$, and $Acc_{i}^{K}$ represents the final accuracy on that task after completing all $K$ tasks.

All models are implemented in PyTorch and trained on NVIDIA GPUs with 24GB of memory. For SER-Diff, the teacher model is first trained on BraTS2020 and subsequently frozen, while the student is updated incrementally using both current-task data and synthetic error maps produced by the teacher. Optimization is performed with the Adam optimizer, an initial learning rate of $1e^{-4}$, and a cosine annealing schedule. A batch size of 4 is employed, and each task is trained for 200 epochs. Standard data augmentation techniques, including random cropping, flipping, and intensity normalization, are applied to enhance model generalization across tasks.

\section{Results}

Table~\ref{tab:results} summarizes the quantitative outcomes on the BraTS2020, BraTS2021, and BraTS2023 datasets. SER-Diff consistently outperforms all baselines across Dice, IoU, and HD95. On BraTS2020, SER-Diff achieves a Dice score of 95.8\%, surpassing U-Net (90.8\%), DMCIE (93.4\%), and EWC (92.1\%). It also reduces boundary error, reaching an HD95 of 4.4 mm compared to 8.1 mm for U-Net, 5.9 mm for DMCIE, and 6.7 mm for EWC. Similar trends are observed on BraTS2021, where SER-Diff improves performance to 94.9\% Dice with an HD95 of 4.7 mm, again outperforming all competing approaches. On BraTS2023, SER-Diff maintains its advantage, achieving 94.6\% Dice and the lowest boundary error of 4.9 mm.  
\begin{table}[ht]
\centering
\caption{Comparison of segmentation performance on BraTS2020, BraTS2021, and BraTS2023 datasets. Metrics include Dice similarity coefficient (Dice), Intersection over Union (IoU), and 95th percentile Hausdorff Distance (HD95, in mm). Best results are highlighted in bold.}
\label{tab:results}
\resizebox{\textwidth}{!}{%
\begin{tabular}{lccccccccc}
\toprule
\multirow{2}{*}{Method} & \multicolumn{3}{c}{BraTS2020} & \multicolumn{3}{c}{BraTS2021} & \multicolumn{3}{c}{BraTS2023} \\
\cmidrule(lr){2-4} \cmidrule(lr){5-7} \cmidrule(lr){8-10}
 & Dice (\%) & IoU (\%) & HD95 & Dice (\%) & IoU (\%) & HD95 & Dice (\%) & IoU (\%) & HD95 \\
\midrule
U-Net \cite{re2}       & 90.8 & 83.2 & 8.1 & 89.7 & 82.4 & 8.5 & 88.9 & 81.5 & 8.6 \\
DMCIE \cite{DMCIE2025}   & 93.4 & 87.9 & 5.9 & 92.5 & 86.8 & 6.2 & 92.0 & 86.3 & 6.1 \\
EWC \cite{kirkpatrick2017overcoming}   & 92.1 & 86.5 & 6.7 & 91.3 & 85.7 & 6.9 & 90.5 & 85.0 & 7.0 \\
SER-Diff (Ours)        & \textbf{95.8} & \textbf{91.2} & \textbf{4.4} & \textbf{94.9} & \textbf{90.4} & \textbf{4.7} & \textbf{94.6} & \textbf{90.1} & \textbf{4.9} \\
\bottomrule
\end{tabular}
}
\end{table}

The improvements in IoU further reinforce these results. SER-Diff reaches 91.2\% IoU on BraTS2020, compared to 83.2\% for U-Net, 87.9\% for DMCIE, and 86.5\% for EWC. On BraTS2021 and BraTS2023, SER-Diff maintains IoU values above 90\%, demonstrating that error-guided replay allows the model to preserve global tumor structure while refining local boundaries. These results indicate that the proposed framework not only mitigates catastrophic forgetting but also enhances overall segmentation consistency across evolving datasets. Beyond numerical metrics, qualitative observations reveal that SER-Diff produces smoother tumor boundaries and recovers small lesion fragments often missed by baseline methods. Visual inspection shows that replayed error maps enable the model to correct subtle discrepancies while maintaining anatomical plausibility. This highlights the strength of synthetic error replay in guiding diffusion-based refinement, ensuring robustness across both coarse- and fine-scale segmentation errors.

\section{Conclusion}
In this work, we introduced SER-Diff, a framework that integrates synthetic error replay with diffusion-based refinement for incremental brain tumor segmentation. By generating and replaying error maps from a frozen teacher, SER-Diff avoids the need for storing raw data while preserving task-specific knowledge across evolving datasets. The diffusion-based refinement process enables precise boundary correction, and the dual-loss strategy balances adaptability to new data with retention of prior knowledge. Experiments on BraTS2020, BraTS2021, and BraTS2023 demonstrate that SER-Diff consistently outperforms strong baselines, achieving higher Dice scores and lower boundary errors while maintaining robustness against catastrophic forgetting. These results highlight the potential of error-guided diffusion models as a promising direction for continual medical image segmentation.

\bibliographystyle{unsrt}  

\bibliography{Ref}     
\end{document}